\documentclass[graybox, nosecnum]{svmult}

\usepackage{mathptmx}       
\usepackage{helvet}         
\usepackage{courier}        
\usepackage{type1cm}        
\usepackage{makeidx}         
\usepackage{graphicx}        
\usepackage{multicol}        
\usepackage[bottom]{footmisc}
\usepackage{hyperref}        
\usepackage{soul}            
\hypersetup{colorlinks=true,urlcolor=blue}
\usepackage[square,numbers]{natbib}
\makeindex             
\graphicspath{ {figs/} }

\newcommand{\fig}[1]{Fig.~\ref{#1}}

\begin{document}
\title*{HERMES-Pathfinder}
\author{Fabrizio Fiore\thanks{Based on work of the HERMES-Pathfinder collaboration, see list in the Appendix}, Alejandro Guzman, Riccardo Campana, Yuri Evangelista }
\institute{
   Fabrizio Fiore \at INAF-Osservatorio Astronomico di Trieste, via G.B. Tiepolo 11, 34143 Trieste Italy,
\email{fabrizio.fiore@inaf.it}
\and Alejandro Guzman \at University of T\"ubingen Institute for Astronomy and Astrophysics, Sand 1, 72076 T\"ubingen Germany,
\email{guzman@astro.uni-tuebingen.de}
\and Riccardo Campana \at INAF-OAS \email{riccardo.campana@inaf.it}
\and Yuri Evangelista \at INAF-IAPS \email{yuri.evangelista@inaf.it}
}

%
%
\maketitle
\abstract{HERMES-Pathfinder is a constellation of six 3U nano-satellites hosting simple but innovative X-ray detectors for determining the positions of, and monitoring cosmic high-energy transients such as gamma-ray bursts and the electromagnetic counterparts of gravitational Wave Events. The HERMES Technological Pathfinder project is funded by the Italian Space Agency, while the HERMES Scientific Pathfinder project is funded by the European Union's Horizon 2020 Research and Innovation Programme under Grant Agreement No. 821896. HERMES-Pathfinder is an in-orbit demonstration, that should be tested in orbit starting in 2023. We present the main scientific goals of HERMES-Pathfinder, as well as a description of the HERMES-Pathfinder payload and performance.}
\section{Keywords} 
Gamma-ray bursts, Nano-satellites, Multi-messenger astrophysics, X-ray all-sky monitor

\section{The advent of multi-messenger astrophysics}

The most dramatic events in the Universe, the death of stars and the coalescence of compact objects to form new black holes, produce some of the most luminous objects in the Universe: gamma-ray bursts (GRBs). However, most of the light is produced far from where the action is, away from the newborn event horizon, the accretion disk, and the region from which a relativistic jet is launched. On the other hand, gravitational waves (GWs), encoding the rapid/relativistic motion of compact objects, give us a direct look into the innermost regions of these systems, providing precise information on space-time dynamics such as mass, spin, inclination, and distance. This information can be greatly enhanced by identifying the context in which the event occurs, which can be done via electromagnetic observations, as the GW/GRB170817 event strikingly showed \cite{Nakar}. While this event hinted at the enormous potential of multi-messenger astrophysics, it has so far remained unique, preventing the full impact of the multi-messenger approach from occurring. The situation will change in the next few years when Advanced LIGO/VIRGO and KAGRA will reach their nominal sensitivity, thus allowing the collection of statistical samples of GW-electromagnetic events with a likely associated electromagnetic counterpart (binary neutron stars (NSs), NS-NS, or black-hole NS systems) in the near future. 

The operation of an efficient X-ray all-sky monitor with good localisation capabilities will have a pivotal role in bringing multi-messenger astrophysics to maturity, and will fully exploit the huge advantages provided by adding a further dimension to our capability to investigate cosmic sources. The HERMES (High Energy Rapid Modular Ensemble of Satellites) project offers a fast-track and affordable complement to more complex and ambitious missions for relatively bright events. The HERMES approach is different from that commonly adopted for space-based observatories: HERMES is an experiment distributed over several nano-satellites. Distributed space architectures are a frontier of innovation in space sciences. They will allow measurements hardly achievable for monolithic systems, such as global and/or high frequency monitoring of cosmic transients. In addition, multi-point observatories would also allow for: a) risk reduction, being naturally more resilient than a single system; b) a drastic reduction in size/mass of the satellites, and consequently cost and development time, especially if the nano-satellite class is considered; c) building the final mission step by step, thus increasing system performance while diluting costs and risks; and d) a full test of the hardware in orbit while preparing for subsequent launches. It is crucial for a project such as HERMES to be developed on a short timescale of a few years compared to one or a few decades for standard space missions 

This gives us a head start into the fascinating new field of multi-messenger astrophysics.
The main requirements for a high energy monitor in the multi-messenger era are 1) The capability to cover the full-sky instantaneously, and 2) localisation capabilities better than a few degrees.
All-sky monitoring is mandatory, because the number of events is, and will remain small, at least for the first and second generations of gravitational-wave interferometers. Therefore missing just one event would mean a primary loss for the science. An all-sky monitor even without good localisation capabilities can address the following key questions: 1) What happens during the merger of compact objects? How frequently does this coincide with short GRBs? How frequent are the formation of powerful relativistic jets? 2) What is the nature of the central engine of short GRBs? What powers the most powerful accelerators of the Universe, neutron star (NS) or black hole (BH) accretion? 3) What is the jet launching mechanism? All these questions can be addressed through simultaneous observations of the gravitational wave event (GWE) and the high-energy emission associated with jet production. Other key questions require the accurate localisation and follow-up of the electromagnetic counterpart of the GWE: 4) Do jets have a universal structure, or does the structure depend on the type of the compact binary? Or would it perhaps depend on parameters such as the mass/spin of the binary components or of the merger remnant? 5) What is the role of compact binary coalescence in the production of heavy elements in the Universe? The jet structure can be addressed through both the analysis of the prompt event and its afterglow, as well as direct VLBI imaging, while element production studies require UV-to-optical-to-NIR spectroscopy of the Kilonova counterpart of the GWE.

HERMES-pathfinder is the first building block of a possible next-generation all-sky monitor. HERMES-Pathfinder is an in-orbit demonstration to show that the architecture and technologies to build a sensitive X-ray and gamma-ray all-sky monitor based on miniaturised instrumentation hosted by nano-satellites are mature \cite{fiore} \cite{sanna}. We present in this paper a description of the the HERMES-Pathfinder payload \cite{evangelista}, including expected performance, and a short description of the space-crafts hosting the detectors. The paper is organised as follows. In Sect. 2, we provide a broad description of the payload developed by the HERMES Technologic and Scientific Pathfinder projects. In Sect. 3, we provide a board description of the HERMES-Pathfinder spacecraft. In Sect. 4, we briefly describe the expected HERMES-Pathfinder performance to illustrate the evolution of the basic concepts to build a powerful observatory in the multi-messenger era.

\section{HERMES-Pathfinder payload}

\subsection{Detector system}

At the core of the HERMES-Pathfinder mission is a hybrid detector concept that is capable of measuring both soft X-rays as well as $\gamma$-rays. The selected design consists of scintillator crystals optically coupled to Silicon Drift Detectors (SDDs). The SDDs can directly detect the soft X-rays up to $\sim$30 keV, while for higher energies the SDDs collect the light produced by the $\gamma$-rays in the scintillator crystals, effectively extending the sensitivity to the MeV range. An exploded view of the detector assembly is shown in \fig{fig:exploded}.  The main components of this assembly are:

\begin{figure}[ht]
\centering
\includegraphics[width=0.95\textwidth]{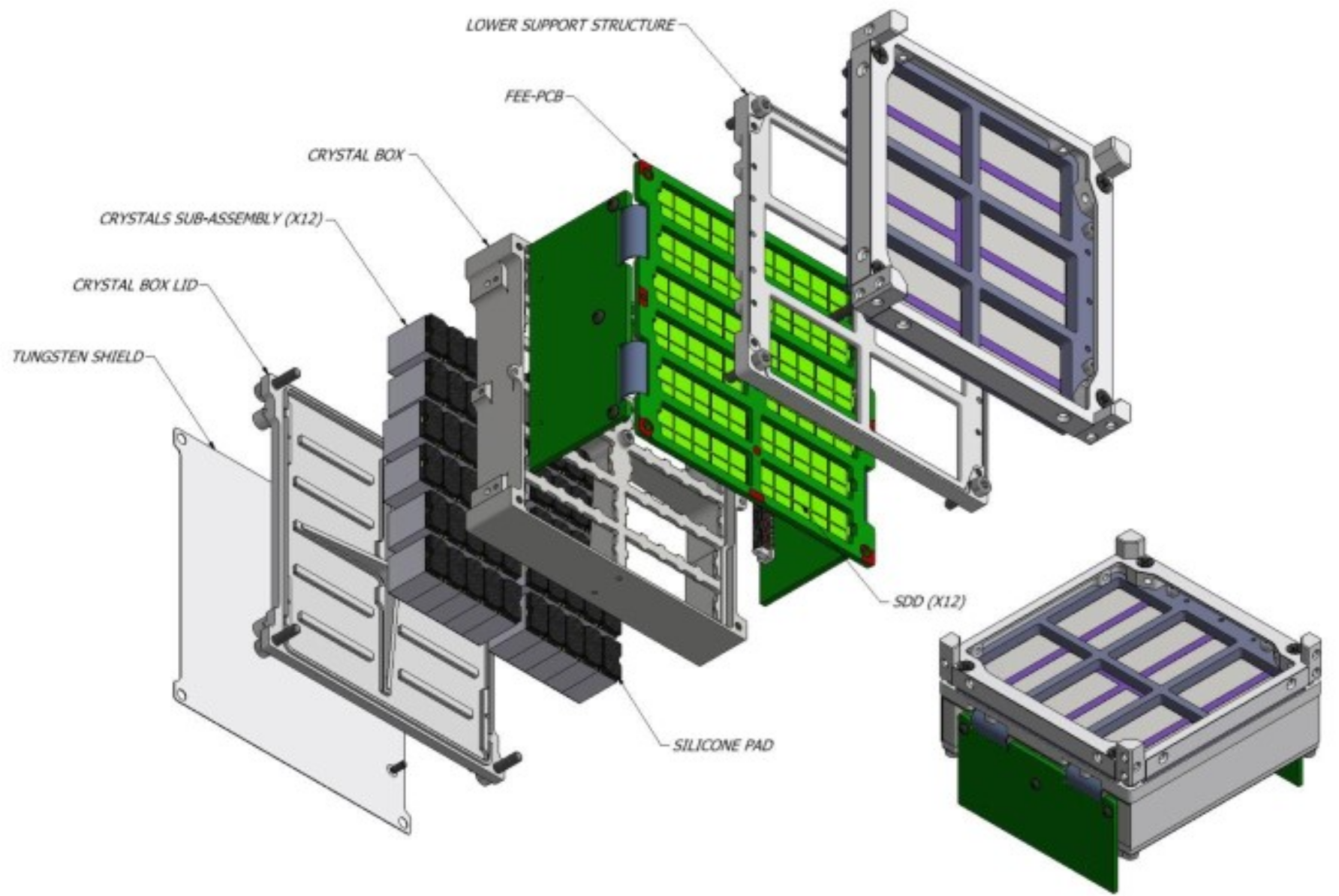}
\caption{Exploded view of the HERMES detector. The detection elements consists of scintillator crystals (grey blocks) optically coupled to the SDDs (green).
The detector support structure, on top of the sensitive plane, allows the mechanical assembly of the detector module, including the optical filters. 
The SDDs and their preamplifiers (not visible) are arranged on the same board (the FEE board), just below the detector support structure. }
\label{fig:exploded}
\end{figure}

\begin{itemize}
 
\item {\bf Optical filter: }
Made out of aluminium deposited on a thin polyimide foil, the optical filter of the HERMES detector is part of the overall optical and thermal design of the P/L. Its prime task is to prevent O/UV light from reaching the SDD detector so as to minimise the current noise generated in the NIR/O/UV band (1130 nm corresponds to the Silicon band-gap). In addition, being the payload thermal design based on passive cooling only, the filter also contributes to the overall thermal design of the detector assembly. 

\item {\bf Detector support structure:}
Made of passivated AISI-316 stainless steel, the detector support structure is composed of  a bottom support structure and a top support structure with the optical 
filter placed in the middle. This assembly represents the mechanical and thermal interface with the payload top rib. The bottom support structure  has an external frame matching the outline of the FEE-PCB, allowing for the required light-tightness and holding purpose.

\item {\bf Multilayer insulation: }
An additional multilayer insulation film (MLI) will be mounted on top of the detector after the final mechanical assembly. This is to ensure a suitable operating temperature for the SDDs.

 \item {\bf Crystal scintillators: }
For the HERMES detectors, the choice of the optimal scintillator material required a careful evaluation of several factors, e.g. the maximisation of the light output (photons per unit of absorbed energy), the non-hygroscopicity of the crystal, its high density and average atomic number (stopping power), the absence of crystal self-radioactivity (background), good radiation-resistance properties, and the light emission characteristic time (as low as possible, to fulfil the scientific objectives). Therefore, the choice has fallen on a relatively recent material, the cerium-doped gadolinium-aluminium-gallium garnet (Ce:Gd$_3$Al$_2$Ga$_3$O$_{12}$ or Ce:GAGG), developed  in Japan around 2010, and commercially available since $\sim$2014 \cite{kamada} \cite{sakano} \cite{yanagida} \cite{yoneyama}.
 
This material has a high intrinsic light output ($\sim$50\,000 ph/MeV), no intrinsic background, no hygroscopicity, a fast radiation decay time of $\sim$90 ns, a high density (6.63 g/cm$^3$), a peak light emission at 520 nm and an effective mean atomic number of 54.4. Each crystal will be coupled to two contiguous SDD channels. Therefore, the effective light output will be split into two channels.
The light output depends on the crystal temperature. For this reason, the effective light output at different temperatures will be calibrated on-ground and the data will be stored in a look-up table for proper in-flight energy calibration using strategically placed temperature sensors. The flight model crystals can be seen in \fig{fig:feepic}.

\item {\bf Silicon Drift Detectors:}
The SDD development builds on  state-of-the-art results achieved within the framework of the Italian ReDSoX collaboration \cite{REDSox}, with the combined design and manufacturing technology coming by a strong synergy between INFN-Trieste and Fondazione Bruno Kessler (FBK, Trento), in which both INFN and FBK co-fund the production of ReDSoX Silicon sensors  \cite{campanadet} \cite{rachevski} \cite{fuschino} \cite{evangelista2} \cite{delmonte1} \cite{delmonte2} \cite{zampa}.

For the development of these SDDss multiple factors and parameters have been optimised to improve the performance of the  HERMES-Pathfinder. Among these parameters are: the pitch and distance of the drift cathodes, the low power consumption of the sensors, a extremely low dark current, high detection efficiency (both for ionising radiation and optical photons), enhanced efficiency on the sensor edges, and a optimised charge collection efficiency.
 
The SDDs are organised in a 2$\times$5 array. As mentioned above, two SDDs are optically coupled to one of the scintillator crystals. The scintillator crystals are optically coupled with the p-side of the SDDs. With this configuration a signal from a single SDD channel will be considered to be originating from the direct absorption of an X-ray in silicon, while two coincident SDD channels (with comparable amplitude) coupled to the same crystal will be considered as a $\gamma$-ray event.
\end{itemize}

\subsection{Electronic boards}

The readout electronics of the detector for  HERMES-pathfinders are in charge of acquiring and digitising the signals produced at the detector SDDs. These are separated into 4 subsystems (`"quadrants'') that work as a redundant ensemble.

\subsubsection{ Front-end electronic (FEE) boards}

\begin{figure}[ht]
\centering
\includegraphics[width=0.45\textwidth]{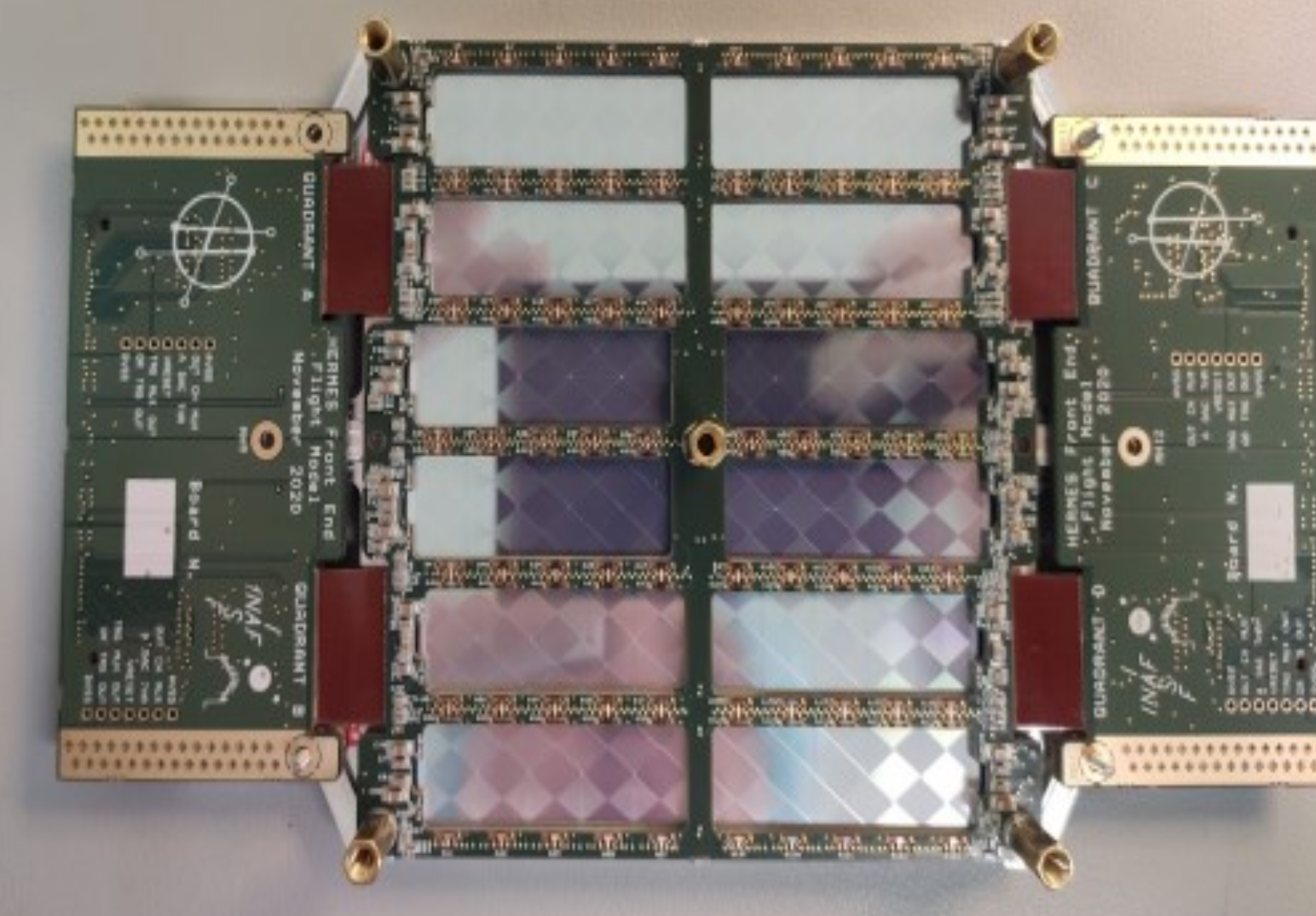}
\quad
\includegraphics[width=0.45\textwidth]{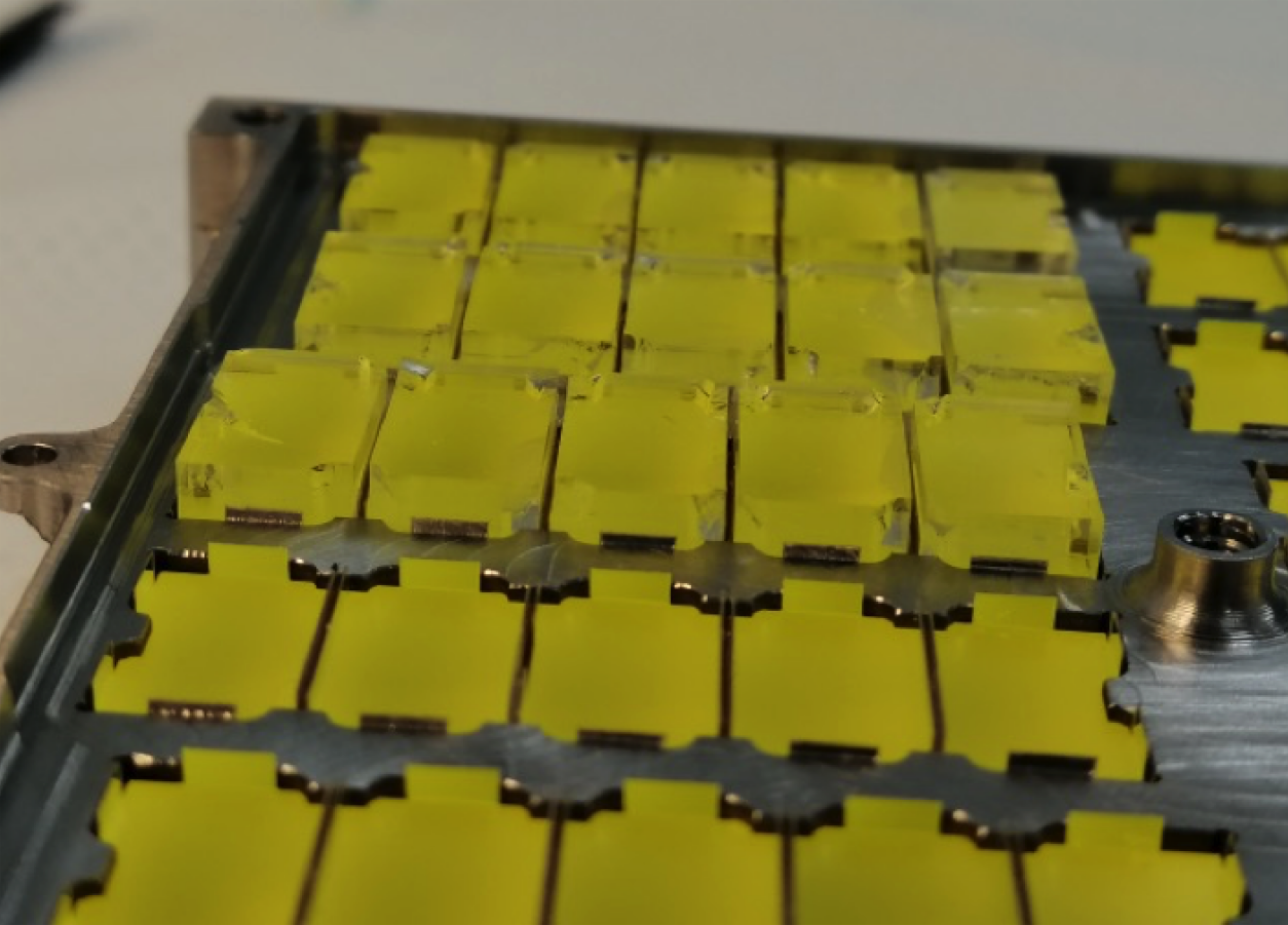}
\caption{ \emph{Left:} Picture of the assembled FEE (without the scintillator crystals). \emph{Right:}A close up of the GAGG:Ce scintillators. }
\label{fig:feepic}
\end{figure}

 The FEE boards are a set of boards accommodating the SDDs and their associated readout \emph{Application Specific Integrated Circuits} or ASICs. The FEE boards accommodate the 12 SDD  matrices, analogue ASICs (see LYRA- ASICs below) and several passive electronic components. The PCB is realised with a Rigid-Flex technology, allowing the 90$^\circ$ bending of the two side wings. Heritage for the analogue FEE ASICs used in HERMES-Pathfinder come from the VEGA ASIC project \cite{ahangarianabhari} \cite{fuschino16}. A photograph of the assembled FEE is shown in Fig.~\ref{fig:feepic}.
 
The analogue readout of the LYRA-ASICs is divided into two sub-components: the LYRA-front end (LYRA-FE) and the LYRA-back end (LYRA-BE).

\begin{figure}[ht]
\centering
\includegraphics[width=0.45\textwidth]{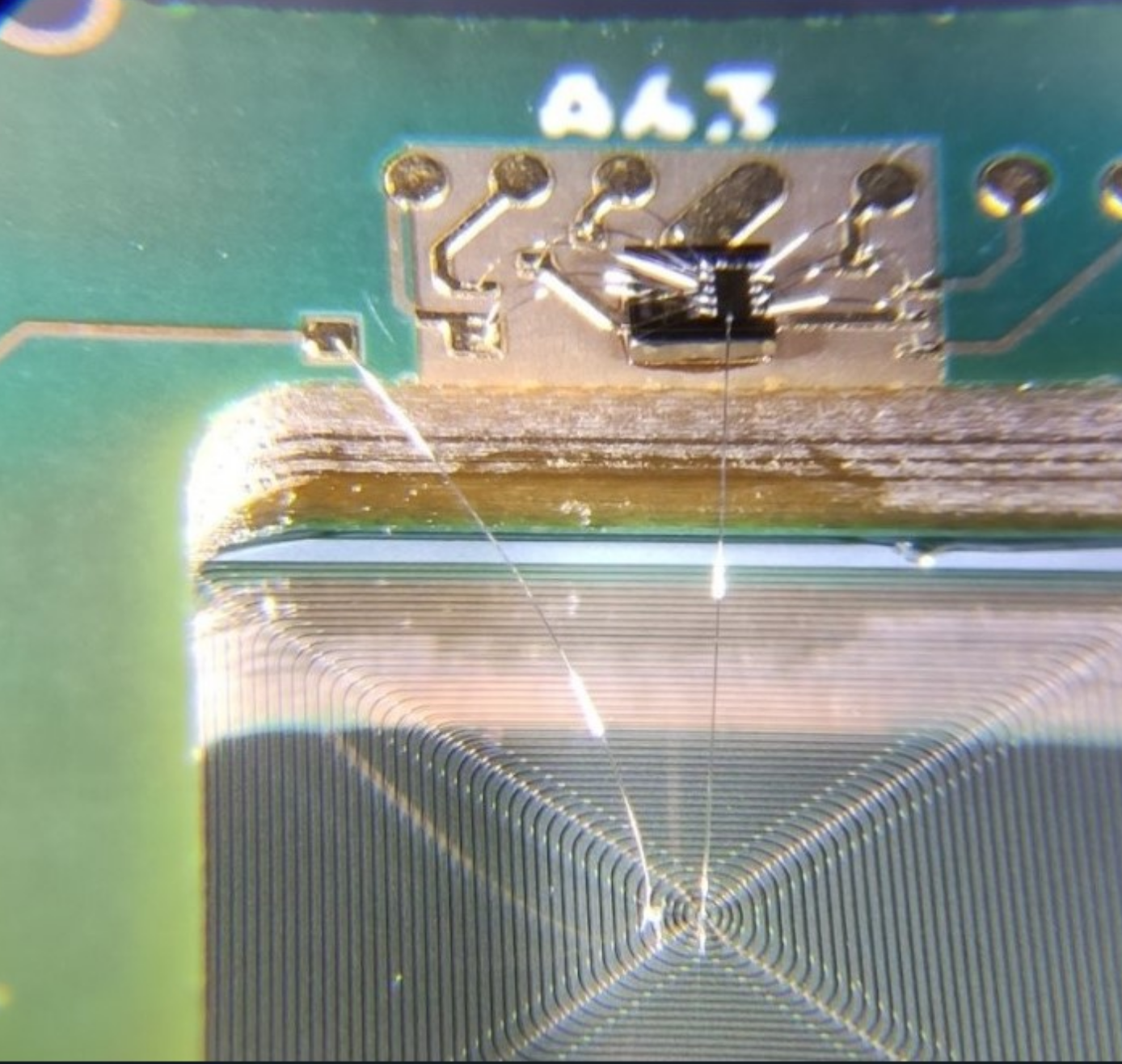}
 \quad
\includegraphics[width=0.45\textwidth]{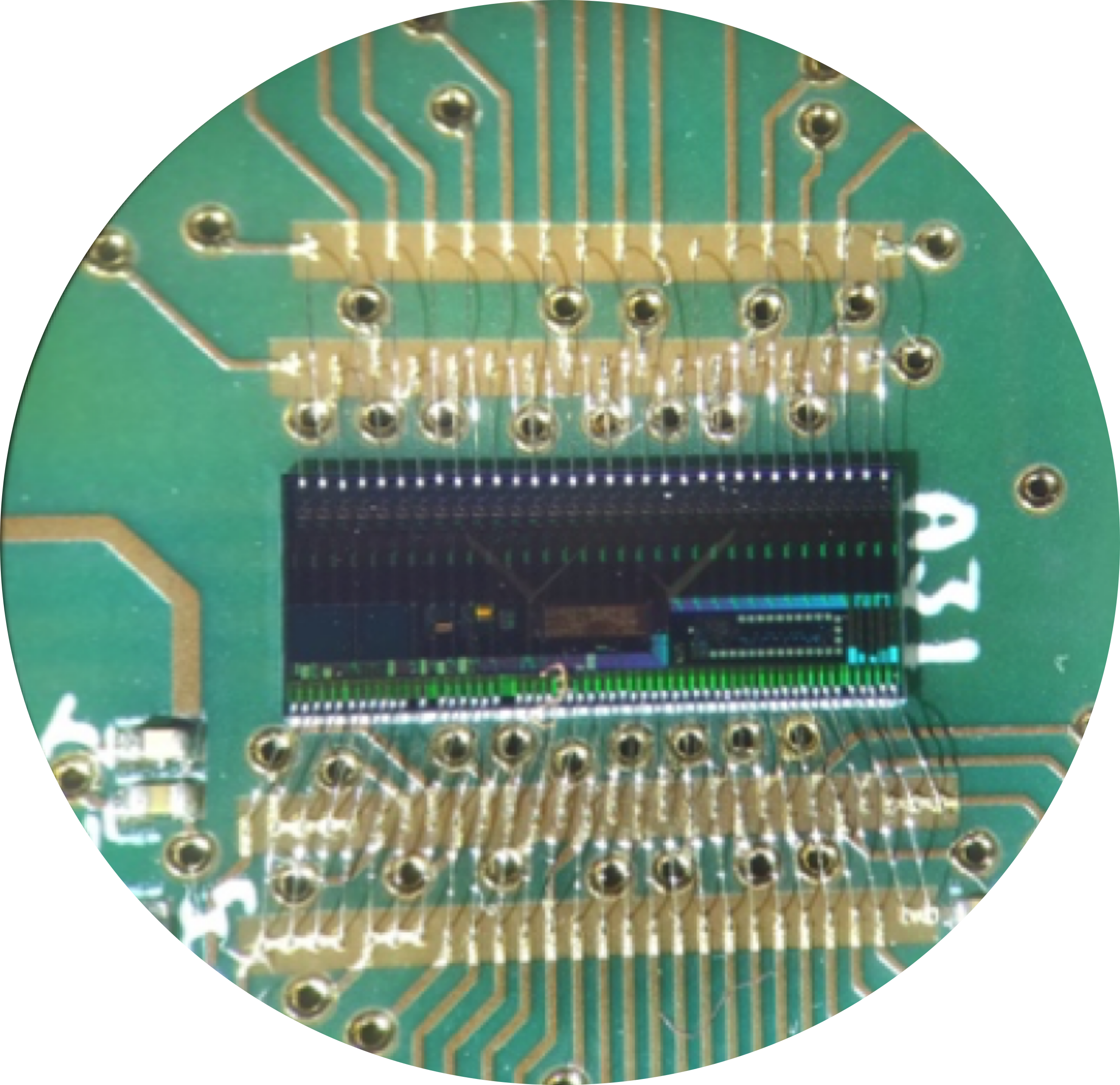}
\caption{Microscope view of the LYRA-FE (left) and LYRA-BE (right) ASICs integrated on the FEE FM.}
\label{fig:lyrapic}
\end{figure}

The {\bf LYRA-FE} chips are arranged on the top of the FEE PCB, which is designed to allow  soft X-rays ($\leq$30 keV) to be detected directly by the SDDs.
For each quadrant of the detector, there are 30 LYRA-FEs coupled to a specific SDD cell. The LYRA-FEs include a pre-amplifier, first shaping stage and signal line-transmitter. These 30 chips are in turn served by one 32-channel {\bf LYRA-BE} (one per quadrant, with two spare channels). The LYRA-BEs complete the LYRA-ASICs signal conditioning by implementing the second shaping stage, discriminators, peak\&hold, control logic, configuration registers and multiplexer. The LYRA-FEs are small chips (0.9$\times$0.6 mm$^2$ die, as shown in \fig{fig:lyrapic} ) to be placed as close as possible to the SDD anodes, in order to minimize the stray capacitance of the detector-preamplifier connection, thus reducing the system noise and maximizing the effective-to-geometric area ratio for the detector plane. The LYRA-BEs chips (6.5$\times$2.5 mm$^2$ die, as shown in \fig{fig:lyrapic}) are still on the FEE board but in a farther position with respect to SDDs, mounted on the two lateral smaller rigid PCBs (a pair of BEs for every side). The detector support structure is placed on the top of this PCB (see \fig{fig:feepic}).

\subsubsection{Back-end electronic (BEE) board}

The BEE is in charge of managing the ASICs configuration, Housekeeping collection, switching ON/OFF of the power lines required by the FEE, manage events data acquisition, transmitting these data to the PDHU, as well as receiving and decoding tele-commands sent by the PDHU. The BEE board also hosts the Chip Scale Atomic Clock (CSAC) which provides the extremely stable 10 MHz clock (synchronised with GPS's  PPS signal) \cite{csac}.

The analogue signals coming from the FEE are routed to the BEE's 16-bit 1 $\mu$s analogue to digital converters (ADCs). There is one ADC per quadrant. Therefore, it is possible to acquire simultaneous events in different quadrants.

The logical core of the BEE is the Intel/Altera Cyclone V FPGA \cite{fpga}. Amongst other features, it was selected for its Single-Event-Latch-Up (SEL) immunity and its low power consumption. This FPGA  implements:

\begin {itemize}
    \item LYRA-ASICs configuration
    \item Event trigger detection
    \item Event acquisition, buffering and transmission
    \item Housekeeping collection
    \item Telecommand parsing and execution
    \item Time management.

\end {itemize}

The BEE communicates with the PDHU through a Serial Peripheral Interface (SPI) bus, with the PDHU as the master on the SPI bus. In this protocol, all data exchanges are initiated by the PDHU (Master of the communication), the BEE receives the command and after correct interpretation executes the function commanded. In a similar manner, every second the PDHU asks  for the acquired data (events/HKs, registers, etc.) present in the BEE and saves it for further processing. To avoid data loss when high event rates occur during observations, the BEE has a  dedicated digital line (a Look-At-Me or LAM signal) that alerts the PDHU to request and read the event buffers immediately.
 
\subsubsection{Power Supply Unit (PSU)}
 
The PSU board is in charge of distributing, regulating, and, if needed, generating all voltages required by the payload. The starting point is the satellite bus, which distributes the following power lines:
\begin{itemize}
\item \textbf{3V3} used to power the BEE, CSAC and the Payload Data Handling Unit (see below).
\item \textbf{5V} powering the linear voltage regulators that provide 3.3 V analogue and digital,and the 2.0 V that powers part of the HK circuitry in the BEE board but also in the detector.
\item \textbf{12V} powering the HV DC/DC converter that provides the operational bias required by the SDD detectors' periphery  ($\approx -123$~V), whereas the anode, in the center, is kept at 2 V.
\end{itemize}

To control these voltages,  electronic switches are placed on the PSU board. However, these switches are controlled by the PDHU, i.e., the PDHU decides which switch must be on or off on the basis of predetermined operational procedures. In addition, another set of switches is commanded by the BEE firmware. This extra layer of security allows the PDHU to command the secondary set of switches via the SPI bus with the BEE, while retaining the BEE's capability to monitor the lines and promptly switch off, in case of anomalous consumption. 

\subsubsection{Payload Data Handling Unit (PDHU)}

The PDHU is the overall controlling system at the payload level, as well as the responsible system for the pre-processing, storage and preparation of the scientific, and housekeeping data.

The selected hardware for the PDHU is the Innovative Solutions In Space (ISIS) On-Board computer (iOBC) \cite{isis}. The iOBC is a high performance and low power processing unit based around a 400 MHz 32bit ARM9 processor. The iOBC was designed specifically for use in Nano-Satellites and has flight heritage since 2012. Particularly important for the HERMES-Pathfinder is the low power consumption (around 400 mW), the redundant 2 GB non-volatile data storage (with industrial-grade SD-Cards), and the 512 kB FRAM which for practical purposes is  impervious to single event upsets.
 All the PDHU interfaces are routed via the PDHU Daughter board directly mated to the iOBC. This DB consists of a shield-design daughterboard which not only interconnects the iOBC with the rest of the payloads subsystems, but also with the satellite bus.
 The PDHU interfaces can be summarised as follows:

 \begin{itemize}
 \item {\bf SBUS interfaces} Used for all communications between the satellite bus and the payload (telemetry and telecommand). They encompass a high speed differential UART and a  I2C connection.
 \item {\bf PSU interface} Includes the 3.3 V power line coming from the PSU and the control lines to the voltage switches present in the BEE.
 \item {\bf BEE interface} Includes the SPI bus, the digital lines coming from the BEE, as well as 7 analogue lines connected to the iOBC's 10-bit ADC converter. These analogue lines are connected to temperature sensors placed in key locations along the payload that are used to calibrate the response of the payload's detector.
 These sensors are very relevant to the HERMES energy calibration since the GAGG:Ce light-output depends on the crystal temperature. For this reason, the effective light output at different temperatures is  calibrated on-ground and the data is  stored in a look-up table for proper in-flight energy calibration.

 \end{itemize}

\subsection{On board firmware and software}
The BEE's firmware is logically subdivided in four independent quadrants, i.e., control of each quadrant runs in parallel inside the BEE's FPGA.   The following entities are included within the firmware :
\begin{itemize}
\item Communication Manager
\item Telecommand Parser
\item Task Execution Manager
\item Housekeepings (HK) Collector 
\item Detectors \& FEE Power Controller
\item FEE ASIC Configuration Manager
\item Trigger Manager
\item Event Acquisition Manager
\item Test Pulse Generator
\end{itemize}
 
The  BEE's firmware is conceptually separated in two states: BEE-Stand-By and  BEE-Data-Acquisition.
On the BEE-Stand-By Mode  it is possible to configure the instrument, loading the configuration tables for ASICs, and acquisition registers. At this operating mode the detector bias High Voltage is OFF but the housekeeping collection can be activated.
These housekeeping blocks contain the actual status of the instrument. From the BEE-Stand-By Mode it is possible to go to the BEE-Data-Acquisition mode.

In the BEE-Data-Acquisition mode the high voltage is ramped up. During this mode, two different acquisitions are possible:  events detected by SDD pixels or test calibration events triggered by the LYRA-FE ASICs stimulated by a dedicated pulse generator commanded by the FPGA. During this mode, the BEE is able to receive triggers from ASICs and manage the complete acquisition of an event, while also producing the housekeeping data and alerting the PDHU, in case the event buffer is 75\% full. This buffer-full-threshold can also be adjusted via a configuration parameter.

\begin{figure}[ht]
\centering
\includegraphics[width=0.95\textwidth]{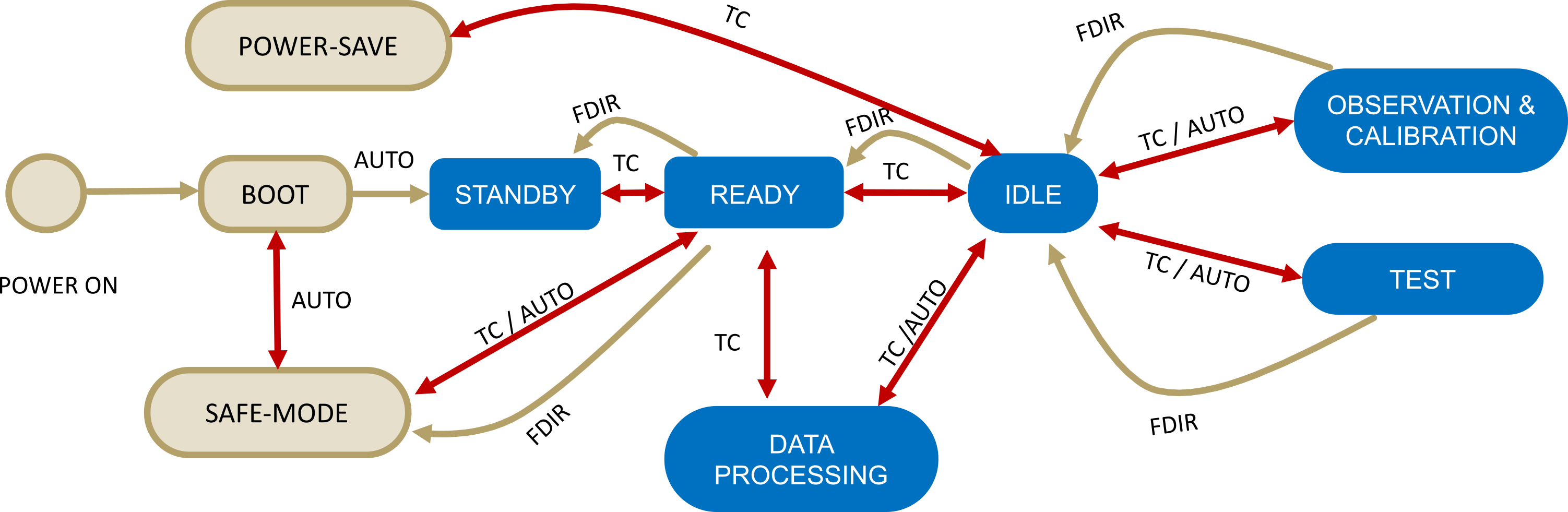}
\caption{A diagram illustrating the interrelation of the payloads operating modes. See text for the explanation.}
\label{fig:opmodes}
\end{figure}

The PDHU's main processor is an Atmel AT91SAM9G20 running freeRTOS. This freeRTOS porting was developed by the iOBC manufacturer and has the same flight heritage as the hardware it is running on. The PDHU's on-board software operates on top of the above-mentioned specifications (microcontroller and OS). Its main functions are instrument control, health monitoring, and science data processing.The PDHU's software design follows a finite-state-machine model that simplifies the operation of the HERMES payload. The different operating modes or  states are (see \fig{fig:opmodes}):

\begin {itemize}
\item {\bf BOOT:} It is the transitory start-up mode at power-on (i.e. when the 3.3 V line goes up).
\item {\bf SAFE-MODE:}. Reserved for exceptional situations as it provides a reduced functionality (only power cycling TC and iOBC's diagnostics). It is triggered via either a TC or when a major error/malfunction has been detected.
\item {\bf STANDBY:}. After power-on the PDHU moves to STANDBY mode without powering the detector. In this state health checks are made, before  proceeding to the READY state.
\item {\bf READY:}. After making health status checks. The PDHU switches ON all power lines and the BEE,  but keeps the detector OFF. All interfaces are available as well as HK data.
\item {\bf IDLE:}. Similar to the READY mode but with the detector powered-on.  Nominally, after observations this is the ``go-to'' operating mode, although it can also be triggered by the Fault Detection, Isolation and Recovery (FDIR) procedure. Data processing (if not too computing intensive) could be carried out in this mode (see below). This mode is also used during the S/C SAA passage, activated by a time scheduled TC.
\item {\bf OBSERVATION AND/OR CALIBRATION:}. From the point of view of the PDHU, both instrument-calibration and scientific-acquisition modes resemble a general OBSERVATION mode whose particular details are configured while in IDLE or STANDBY mode. It is foreseen to initiate this mode with a TC, although it could also be scheduled.
\item {\bf TEST:}. This mode is introduced to implement a merge of IDLE and OBSERVATION modes. Its main purpose is to perform specific checks for diagnosis purposes. This mode is also only to be triggered via TC. Some of the foreseen tests include test(the following list is not exhaustive): to check the ADC-readout, to check the SPI-communication channel, a BEE readout test, a CSAC test, (see below for description of these interfaces).
\item {\bf DATA PROCESSING:}. Its main purpose is to prepare the scientific data packets for transmission to ground. This includes on-board burst-searching algorithms, data reducing/compressing, and the collection of the housekeeping reports. Although this operational mode is conceived as a stand-alone mode, running some of its tasks while in OBSERVATION mode is also a possibility. This would allow for a prompt acknowledgement of scientifically interesting events.
\item {\bf POWER SAVE:}. This mode is introduced to implement a low power consumption mode. Its main purpose is to lower the HERMES Payload power consumption to cope with a time period of reduced power availability from the sBUS, still maintaining the Payload switched on and monitoring the Payload health status. This mode is triggered via either TC or a time-scheduled change in the operation mode.
\item {\bf FDIR:}. Although not an operating mode on its own, the Fault Detection Isolation and Recovery (FDIR) approach of the Payload is model-based, and relies on constant monitoring of the on-board sensors (temperatures, currents and voltages). These readings are compared with an on-board database of updateable nominal ranges. If values are out of the expected ranges the FDI tasks emits an alert and assess the severity of the fault. Only in high severity cases an automatic power-cycling of the Payload is triggered. This database is kept in the PDHU's FRAM for extra security.

\end {itemize}

\subsubsection{Data handling }
Due to data downlink constraints, the PDHU needs to  reduce the acquired events before   transmitting them  to ground. However, a fast-alerting system  is crucial  to allow for follow up observations of GRBs. The procedure commences by differentiating between events originating solely from the SDD (labeled X-ray events) and the ones originating with the interaction of gamma-rays and the GAGGe crystals.  For X-ray events only  SDD is activated, whereas for the gamma-ray events at least two SDDs are activated simultaneously (share the same time stamp). The number of triggered SDDs is  reported by the BEE and readout by the PDHU. The temporal check is carried out by the PDHU and together with the ancillary data (channel specific gain, temperature, and diverse HK data) the energy is assigned using  a calibration Look-Up-Tables (LUT), that take into account all this ancillary data. This will allow the possibility  to downlink all the relevant transient phenomena (GRBs, solar flares or other fast transients).
 
 In addition to the transient phenomena data the PDHU collects scientific ratemeters that include count-rates for different time scales ranging from milliseconds to several seconds (e.g. 0.001, 0.016, 0.064, 0.256, 1, 10, 100 seconds). These are also  separated  by energy range and the geometrical region of the detector (quadrant).

The triggering scheme is a multi-tier and  multi-channel scheme. It includes the following levels:
\begin {itemize}
\item {\bf Level 1 trigger:} implemented at the BEE level. It is activated if the ADC-counts of a given channel are above an adjustable threshold. This threshold is set with an 8 bit DAC assigned.
\item {\bf Level 2 triggers (burst search algorithms):} The transient-search routines (referred to as "burst search" routine) find a transient above an adjustable likelihood value. These triggers are applied at specific energy bands (three different energy bands) and for different time windows \cite{pdhuspie}. This Burst Search (BS) algorithms rely on two approaches. The first approach is done by the PDHU using payload ratemeters. Based on different ratemeters, it will integrate data in different energy bands, timescales and geometric regions. Since the GRBs are strongly energy and timescale dependent, different results for the different ratemeters are expected. The threshold energy defining the three energy channels are re-programmable on flight, as well as the integrating timescales. The second approach for transient search follows a Bayesian approach \cite{belanger} assuming a Poisson distribution for the background.

\end {itemize}

To further exploit the novel concept of the HERMES-Pathfinder, its satellites will have the ability to emit a real-time burst alert, relying in the IRIDIUM satellite constellation. For the most scientifically interesting observations, the burst alerts are accompanied with ``ready to use'' data. They can be made available to the user much earlier than regular data as they would exploit the available telemetry budget with the IRIDIUM constellation. 
This is called the HERMES  Quick-Look Data (QLD), and although not in the baseline data products for the HERMES-Pathfinder (not required to fulfill the scientific objectives of the mission), it is highly desirable for increasing the scientific output of the mission and follow-up observations. This QLD compromises a 340 byte package  transmitted immediately to ground containing: the satellite identification number, GPS information (including location at the time of the triggered), three light curves at different energy bands containing the count-rates before, during, and after the alert was issued, and ancillary data pertaining to the health and overall status of the detectors (count rates, status of other triggers, temperatures, currents, etc.)

\section{HERMES-Pathfinder Service Module}

The Service Module (SM) used in the HERMES-Pathfinder project is a CubeSat of the 3U class. The SM can be broken down to the following on-board sub-systems: Structure \& mechanisms; Electric Power (EPS), thermal Control (TCS), Telemetry, Tracking \& Command (TTC), Attitude and Orbit Determination and Control (AODCS) and On Board Data Handling (OBDH). The main design driver was to ensure, at the minimum, the space segment basic functionalities whenever non-nominal conditions occur, to be robust to mortality, typically occurring because of lack of electronics: the survive and communicate functionalities are mandatory even before science operations accomplishment. Therefore, the EPS and TTC sub-systems drove the design. Because of a continuously operating detector, even in Standby Mode, and the need to be continuously ground-connected for scientific events occurrence communication, the electric power resource may represent a criticality: CubeSat in fact stays in the order of a few watts supply, because of the limited surface to body mount solar cells. To increase the available power to 20--30 W, two deployable wings are included in the SM. The TCS will be passive at the most, to save power request on board; TTC will focus on the science data robust transmission, taking advantage of the IRIDIUM module to ensure RF connection anytime a GRB event may occur; downloading of scientific data will be done using an S-band transceiver/antenna system. To ensure robustness, omnidirectional antennas are mounted on board supporting UHF/VHF connections, to protect the mission loss from attitude control potential anomalies. No orbital control is present, while the state vector determination will benefit of the GPS receiver and the inertial measurement unit (IMU) for the centre of mass, and magnetometers and sun sensors for the attitude; attitude will be controlled, using magneto-torquers and reaction wheels for the platform health and operability maintenance.

\section {HERMES-Pathfinder performance}

The main drivers for the HERMES-Pathfinder performance are the effective area, the timing resolution, and the expected in-orbit background. All these contributions have been evaluated by means of dedicated simulations and measurements.

\begin{figure}[ht]
\centering
\includegraphics[width=0.7\textwidth]{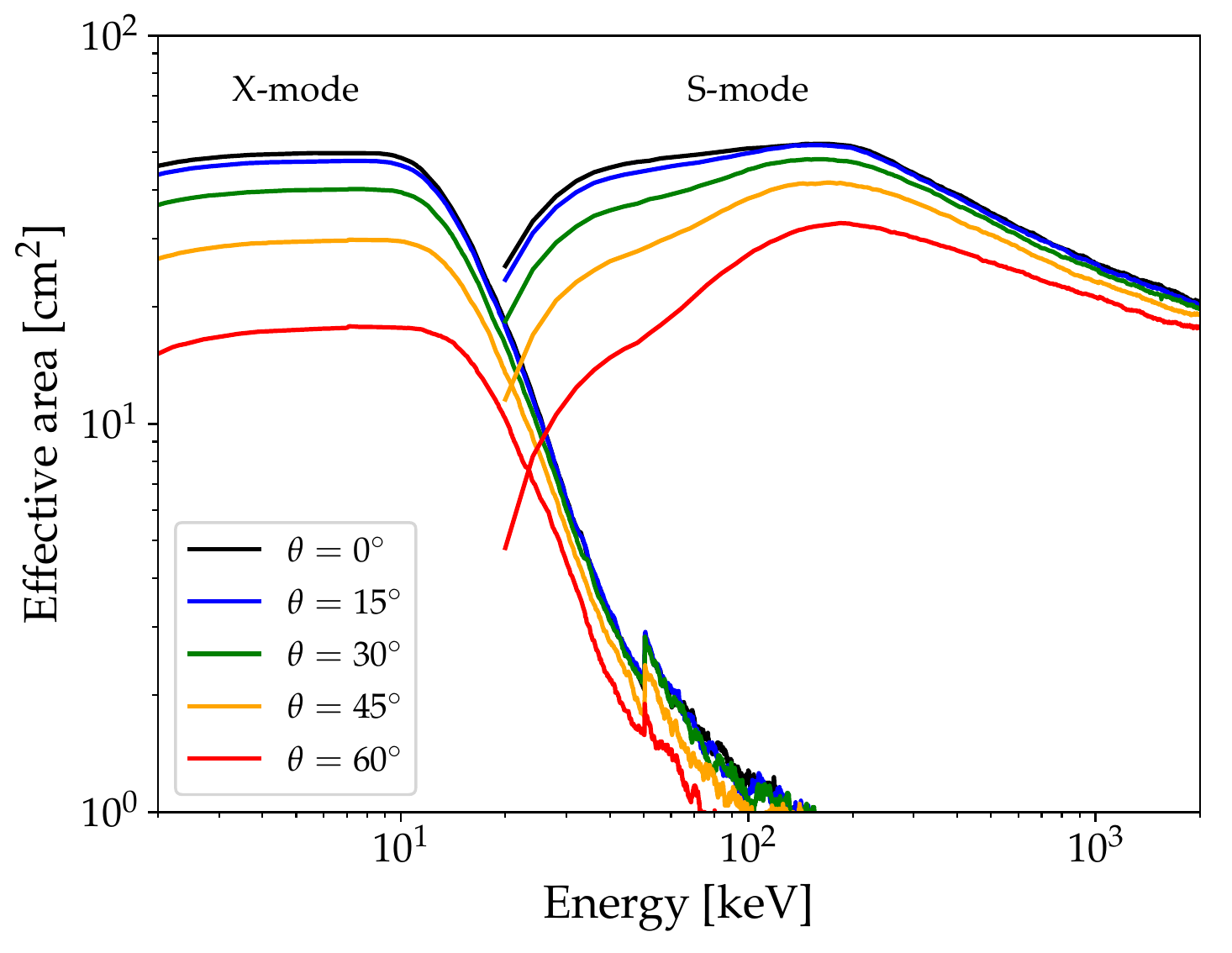}
\caption{The HERMES-Pathfinder effective area, as a function of the off-axis angle.}
\label{fig:effarea}
\end{figure}

The detector effective area has been estimated using analytical and Monte Carlo calculations taking into account the geometric area of both the SDDs and the crystals, the transmission of the thermal blanket and of the optical filter, the SDD quantum efficiency for X-ray detection (X-mode), and  the GAGG efficiency for $\gamma$-ray detection (S-mode), the latter also  involving  the efficiency of the optical contact between SDDs and scintillators. Figure~\ref{fig:effarea} shows the overall effective area for X-mode (black) and S-mode (blue), also as a function of the source off-axis angle. The decrease at low energies for X-mode is due to the blocking effect of all the passive layers in front of the detector, while the effect of decreasing silicon efficiency is apparent above $\sim$10 keV. 

The time resolution (i.e., the error in the determination of a photon arrival time with respect to a local, on-board frame of reference) and time accuracy (i.e., the error in the synchronisation of the local time reference frame with respect to the Universal Coordinated Time) for the HERMES-Pathfinder payload have been estimated taking into account the GPS/PPS contribution and the physics of the photon detection in both X and S-modes. For the X-mode, the total time tagging precision is estimated to be around 320 ns (with GPS locked) increasing to 370 ns (with GPS unlocked). For S-mode, we have about 220 ns and 280 ns, respectively. In summary, the overall photon time tagging precision is thus expected to be lower than 300--400 ns, according to the operative mode.

Instrument dead time is constrained by the time required by the BEE for the analogue-to-digital conversion of the event. By design, the conversion time is $\leq$15~$\mu$s, typically 12~$\mu$s, and is fixed. Therefore, up to a maximum source count rate of 4$\times$10$^4$ counts/s the dead time is below 10\%.

The main contributions for the scientific HERMES-Pathfinder background count rate are due to the following sources, found in the foreseen operating low-Earth equatorial orbit:
\begin{enumerate}
\item Cosmic X-ray diffuse background
\item Earth gamma-ray albedo emission
\item Earth neutron albedo emission
\item Primary cosmic rays (protons, leptons, alpha particles)
\item Secondary cosmic rays atmospheric emission (protons, leptons)
\item Activation of spacecraft structure due to high-energy particle environment
\end{enumerate}

\begin{figure}[ht]
\centering
\includegraphics[width=0.7\textwidth]{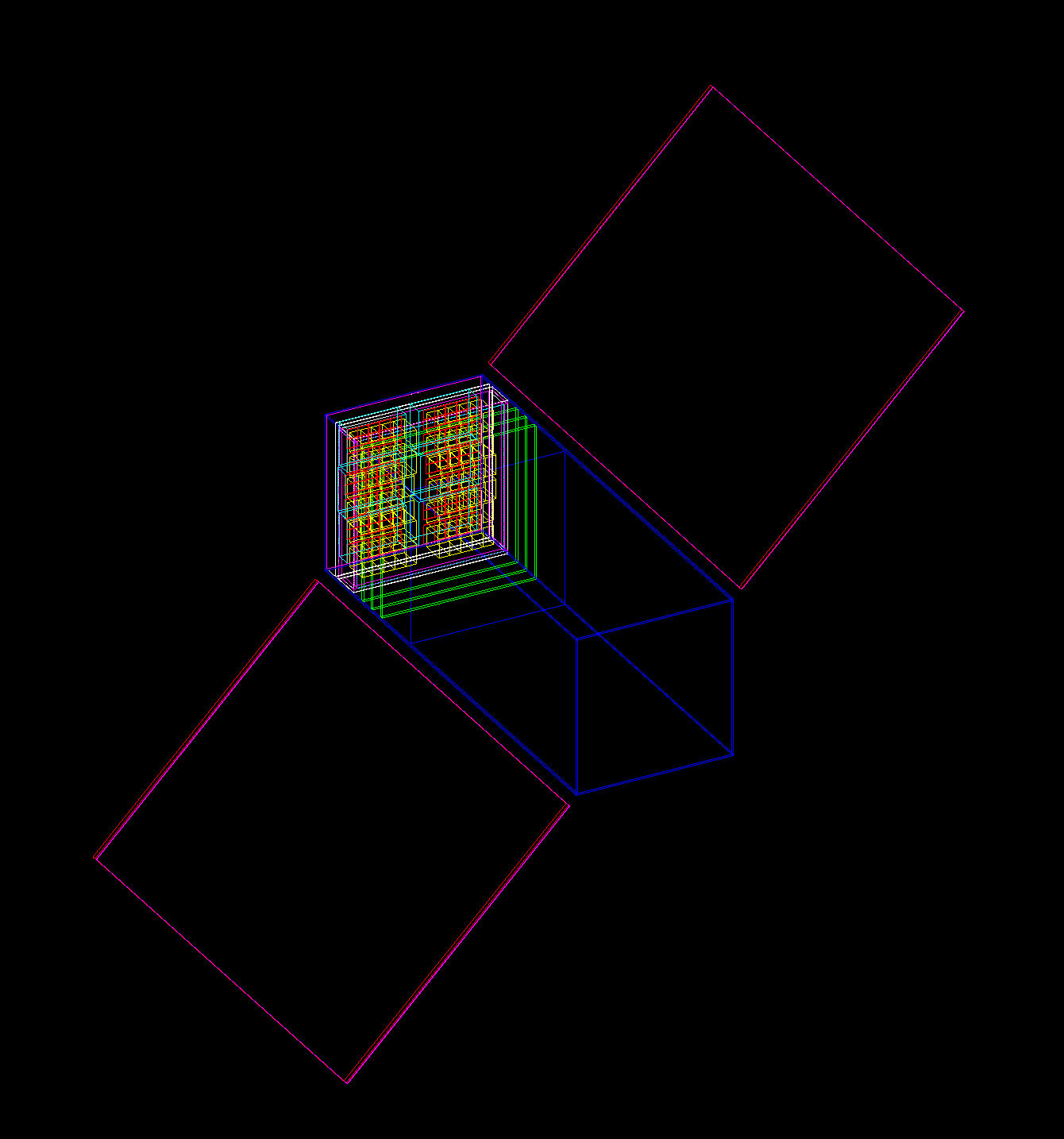}
\caption{The HERMES-Pathfinder Geant-4 mass model.}
\label{fig:massmodel}
\end{figure}

\begin{figure}[ht]
\centering
\includegraphics[width=0.7\textwidth]{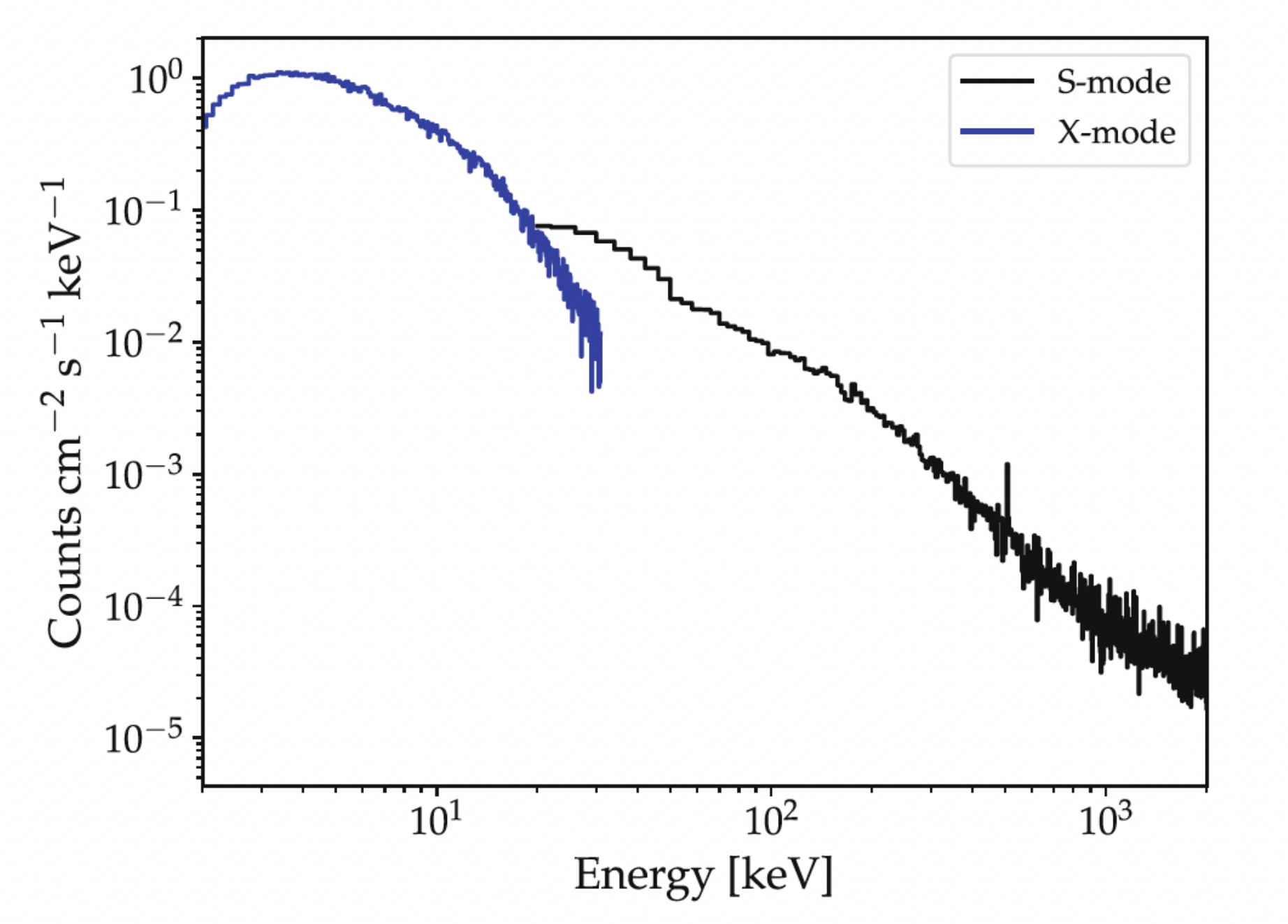}
\caption{The HERMES-Pathfinder expected background.}
\label{fig:bkg}
\end{figure}

These sources have been used as input for the background evaluations by means of Monte Carlo Geant-4 mass model simulations \cite{campana20}, \cite{campana13}. A Geant-4 mass model of the payload has been developed. A simplified geometrical and physical description of the detector structure and materials has been implemented, and the interaction of photons and particles with the mass model has been simulated, in order to derive the detector response matrices (redistribution matrix and effective area as a function of energy) and the expected scientific background. In Fig.~\ref{fig:massmodel} the mass model as implemented in Geant-4 is shown, while Fig.~\ref{fig:bkg}  reports the resulting scientific background.

\section{Programmatic status}

The integration and test of HERMES-Pathfinder proto-flight model (PFM) started on August 2021 with the integration of the detector system at Fondazione Bruno Kessler (FBK) laboratories in Trento. The detector system PFM underwent functional tests at both FBK laboratories and INAF-IAPS laboratories in Rome during Q3/Q4 2021. The integration and test of the full payload PFM (detector system and electronic boards) started in March 2022 at INAF-IAPS laboratories. A careful on ground calibration, including measurements of the 120 sensors gain as a function of the temperature will follow in the next weeks. Integration and test of the SM PFM started on February 2022 at Politecnico di Milano Laboratories. Integration of the payload PFM in the SM is scheduled for May 2022. The full PFM qualification review is scheduled for June 2022. 

Integration and test of the payload and SM for the other 5 flight models (FMs) started in March 2022 and will end on Q4/2022. The acceptance review of the six FM is scheduled for the end of 2022. The constellation should be deployed in orbit during 2023.

HERMES-Pathfinder will be served by two dedicated and identical ground stations, one in the ASI centre at Malindi, Kenya, and the second in the Katherine site, Northern Territories, Australia. The Malindi station is owned by ASI and will be operated under ASI responsibility, the Katherine station is owned by a consortium including the INAF, the University of Masaryk, the University of Tasmania and the University of Melbourne. Support for the Katherine station includes funds from the H2020 INFRAIA project AHEAD2020, GA n. 871158.

\textbf{Acknowledgements}
The authors acknowledge support from the Horizon 2020 Research and Innovation Programme under Grant Agreement No. 821896 HERMES-Pathfinder, the Horizon 2020 INFRAIA Programme under Grant Agreement n. 871158 AHEAD2020, and  the ASI-INAF ``Accordo Attuativo'' HERMES Technologic Pathfinder n. 2018-10-H.1-2020.

\appendix

The HERMES-Pathfinder collaboration includes the following researchers and institutes:

F.~Fiore$^a$, L.~Burderi$^b$, M.~Lavagna$^c$, Y.~Evangelista$^a$, R.~Campana$^a$, F.~Fuschino$^a$, A.~Sanna$^b$, P.~Lunghi$^c$, A.~Monge$^e$, R.~Bertacin$^d$
% ...

$^a$ INAF via del Parco Mellini 84, I00136 Roma, Italy; 
$^b$ Dipartimento di Fisica Università degli Studi di Cagliari, Italy; 
$^c$ Dipartimento di Scienza e Tecnologia Aerospaziali, Politecnico di Milano, Italy; 
$^d$ Agenzia Spaziale Italiana, Italy; 
$^e$ DEIMOS, Spain; 
% ...

\end{document}